\documentclass[final,5p,times,twocolumn]{elsarticle}

\usepackage{amssymb}
\usepackage{tabu,multirow}

\newcounter{bla}

\journal{Computer Physics Communications}

\begin{document}

\begin{frontmatter}

  \title{MatchingTools: a Python library for symbolic effective field theory
    calculations}

  \author[a]{Juan C. Criado \corref{author}}

  \cortext[author] {Corresponding author.
    \\\textit{E-mail address:} jccriadoalamo@ugr.es}
  \address[a]{CAFPE and Departamento de F\'isica Te\'orica y del Cosmos,
    Universidad de Granada, E-18071 Granada, Spain}

  \begin{abstract}
    \emph{MatchingTools} is a Python library for doing symbolic calculations
    in effective field theory. It provides the tools to construct general
    models by defining their field content and their interaction Lagrangian.
    Once a model is given, the heavy particles can be integrated out at the
    tree level to obtain an effective Lagrangian in which only the light
    particles appear. After integration, some of the terms of the resulting
    Lagrangian might not be independent. \emph{MatchingTools} contains
    functions for transforming these terms to rewrite them in terms of any
    chosen set of operators.
  \end{abstract}


\end{frontmatter}

{\bf PROGRAM SUMMARY}

\begin{small}
  \noindent
  {\em Program Title: MatchingTools} \\
  {\em Licensing provisions: MIT} \\
  {\em Programming language: Python (compatible with versions 2 and
    3)} \\
  {\em Nature of problem:} \\
  The program does two kinds of calculations: computing an effective
  Lagrangian for the light fields of a field theory by integrating out at the
  tree level the heavy fields and performing algebraic  manipulations with
  tensors in the (effective) Lagrangian. \\
  {\em Solution method:}\\
  The tree level integration of heavy fields is done by substituting them
  inside the Lagrangian by a covariant derivative expansion of the solution to
  their equations of motion. The transformation of Lagrangians is implemented
  as an algorithm for finding patterns of tensor products and replacing them
  by sums of other products.
\end{small}

\section{Introduction}
\label{intro}

When studying physical phenomena in the framework of field theory, it is often
convenient to describe the low energy behavior of the system in a way that
does not involve the heavy degrees of freedom. A low energy effective theory
can be derived from a more fundamental one, when the latter is known. The
connection between both descriptions is done by integrating out the heavy
fields. The basic idea is to find the set the effective interactions of the
light fields such that the corresponding low energy observables match, to the
desired precision, those computed using the full theory
\cite{Georgi:1985kw, Donoghue:1992dd}.

An important example arises in particle physics, when studying extensions of
the Standard Model. The latest experimental results from the Large Hadron
Collider (LHC) do not show any evidence of direct production of new particles
(see for example
\cite{Aaboud:2016qeg, Aaboud:2017qpr, Sirunyan:2017ukk, Sirunyan:2017wto}).
Therefore, the discovery of new physics arising within the range of energies
of the current phase of the LHC seems more and more unlikely.

In view of this perspective, it is interesting to extract features of physics
at higher, currently unreachable energies by using precision measurements.
This can be done using an effective theory approach. In the Standard Model
Effective Field Theory (SMEFT), the Standard Model is extended to include
non-renormalizable operators (see \cite{Brivio:2017vri} and references
therein). In this setting, the high energy physics is parametrized in low
energy by the coefficients of the new operators. These coefficients can be
constrained by the experimental data
\cite{
  Han:2004az, deBlas:2013qqa, deBlas:2013gla, Gupta:2014rxa, Ellis:2014jta,
  Falkowski:2014tna, deBlas:2015aea, Berthier:2015gja, Pomarol:2013zra,
  deBlas:2017wmn}.
Another effective approach to extending the Standard Model is the Higgs
effective theory, in which the gauge symmetry is realized non-linearly
\cite{
  Appelquist:1980vg, Longhitano:1980iz, Longhitano:1980tm, Feruglio:1992wf}.

Many of the proposed theories for physics beyond the Standard Model predict
the existence of new, heavy particles
\cite{Georgi:1974sy, Fritzsch:1980iu, ArkaniHamed:1998rs, Randall:1999ee}
The result of integrating out these heavy fields is the collection of the
corresponding coefficients of the operators of the SMEFT
\cite{
  delAguila:2000rc, delAguila:2008pw, delAguila:2010mx, deBlas:2014mba}.

The procedure of matching can be described algebraically in terms of tensor
calculus manipulations involving the computation of functional derivatives and
the substitution of heavy fields by other previously obtained expressions
\cite{
  Fraser:1984zb, Aitchison:1984ys, Aitchison:1985pp, Aitchison:1985hu,
  Chan:1985ny, Chan:1986jq, Gaillard:1985uh, Cheyette:1985ue,
  delAguila:2016zcb, Ellis:2016enq, Fuentes-Martin:2016uol, Zhang:2016pja}.
The complexity of the process quickly grows with the number of heavy fields
and their interactions. It is in this context where the development of a
computer tool to automatize the process becomes necessary.

\emph{MatchingTools} can perform tree-level integration of heavy fields in any
given Lagrangian. It has been developed with the application to the SMEFT in
mind, but it is able to work with any situation describable by a Lorentz
invariant field theory in which the high energy degrees of freedom to be
removed are scalars, vector-like or Majorana fermions, or vectors. By
introducing the generic solution to their equations of motion, other types of
fields can be treated as well. The validity of \emph{MatchingTools} extends to
any level in the expansion in inverse powers of the cut-off energy of the
effective theory.

The Lagrangian resulting from integration usually contains redundancies:
operators that can be written in terms of others using identities of the
symmetry group, integration by parts or equations of motion of the light
fields
\cite{Kallosh:1972ap, Politzer:1980me, Georgi:1991ch, Arzt:1993gz}.
A complete set of operators that are independent under this set of
transformations is called a basis. Several such bases have been described
\cite{Grzadkowski:2010es, Contino:2013kra, Masso:2014xra}.

The transformation of the results of an integration to a chosen basis can
also be done using \emph{MatchingTools}. One should introduce the identities
between tensor expressions needed to transform some operators into others, as
well as the desired basis.

There are other tools for the manipulation of bases of operators, such as
\emph{Rosetta} \cite{Falkowski:2015wza}. The portion of \emph{MatchingTools}
that deals with this calculations differs from it in two main points: first,
it allows not only for the transformations between sets of already independent
operators, but for the transformation of any set of operators into a basis.
Moreover, \emph{MatchingTools} has the ability of doing transformations not
with the operators themselves, but with parts of them, allowing for general
transformations between parts of tensor expressions into others. Actually,
\emph{MatchingTools} can be used as system for tensor calculus manipulations,
not necessarily in the context of an effective field theory. It provides a
fast way of doing complex symbolic calculations with many fields and terms
involved, which is safe against algebraic errors.

A direct application of \emph{MatchingTools}, which has also served as an
extensive check of its validity, is the integration of all possible new fields
that have linear gauge-invariant renormalizable couplings to the Standard
Model, keeping terms up to dimension six in the results \cite{mixed}.

A package that implements a similar way of dealing with the specification of
models is \emph{FeynRules} \cite{Christensen:2008py, Alloul:2013bka}, thought
its objectives are completely distinct. One possible direction for future work
with \emph{MatchingTools} is making the connection with \emph{FeynRules}.

Among other computer tools for calculations in the context of the SMEFT we
have \emph{DsixTools} \cite{Celis:2017hod} (which allows for several
calculations including a case of tree level matching) and \emph{SMEFTsim}
\cite{Brivio:2017btx} (which is able to produce theoretical predictions and
constraints for the Wilson coefficients of the dimension 6 SMEFT).

\emph{MatchingTools} is available in GitHub
(https://github.com/jccriado/matchingtools) and in the PyPI repository
(https://pypi.python.org/pypi/matchingtools/), so it can be installed using
pip \cite{pip} as

\begin{verbatim}
    pip install matchingtools
\end{verbatim}

This article is organized as follows: section \ref{sec:theory} describes the
procedure and the formulas that the library uses for the integration of heavy
particles. Sections \ref{sec:models}, \ref{sec:integration},
\ref{sec:transformations} and \ref{sec:output} explain the features of
\emph{MatchingTools} and how to use it. Section \ref{sec:example} proposes a
simple example that serves to see the library in action and as a test case.
Some extra features for the applications in physics beyond the Standard Model
are introduced in section \ref{sec:extras}. Section \ref{sec:others} is an
explanation of how to integrate out new types of fields that are not included
in \emph{MatchingTools}.

\section{Theoretical framework}
\label{sec:theory}

\subsection{Tree level integration}

Starting with a high energy theory with action $S[\phi,\Phi]$ depending on the
light and heavy fields $\phi$ and $\Phi$ the effective action $S_{eff}$ for the
light fields is obtained through:
\begin{equation}
  e^{iS_{eff}[\phi]} \propto \int \mathcal{D}\Phi e^{i S[\phi,\Phi]},
\end{equation}
where $\int\mathcal{D}\Phi$ means integrating over all the configurations of
the heavy fields $\Phi$. The configuration that contributes the most to this
integral is the classical configuration $\Phi_c$, which extremizes the action.
To leading order in $(\Phi-\Phi_c)$, we get
\begin{equation}
  S_{eff}[\phi] = S[\phi,\Phi_c],
\end{equation}
known as the tree level approximation. It is the one that we will use in this
article.

\subsection{Equations of motion and their solution}

To obtain the classical configuration of the heavy fields it is necessary to
solve their equations of motion. They are determined from the condition:
$\delta S/\delta \Phi = 0$.

The variation of a local action $S=\int d^mx \mathcal{L}$ can be written as
\begin{equation}
  \frac{\delta S}{\delta \Phi} =
  \sum_n(-1)^nD_{\mu_1} D_{\mu_2} \cdots D_{\mu_n}
  \frac{\partial \mathcal{L}}
  {\partial (D_{\mu_n}\cdots D_{\mu_2}D_{\mu_1} \Phi)},
\end{equation}
which we have expressed in terms of a covariant derivative $D$ for the gauge
group of the low energy effective field theory. It is convenient to split the
action into a quadratic and an interaction part:
\begin{equation}
  S = \int d^m x \mathcal{L}_{quad} + S_{int},\qquad
  \mathcal{L}_{quad} = -\Phi^\dagger P \Phi,
\end{equation}
where $P$ is some differential operator.

For the bosonic fields, the solution to the equation of motion will be given
by the application of the inverse of $P$ to a functional derivative of the
interaction action. $P^{-1}$ can be expanded in each case in powers of
$D_\mu / M$. For fermions, the solution will be given as a system of two
equations. Recursive substitution of one into the other will give the solution
to any order in $D_\mu/M$. Because we usually limit the dimension of the
operators appearing in the effective Lagrangian we will only need to
substitute a finite number of terms of these infinite expansions.

Several fields can be integrated out together. The solution to the equation of
motion of each of them may involve the others, but they can be replaced
recursively by their corresponding solutions to the equations motion to obtain
solutions that only involve the light fields to the desired order.

The Lagrangian $\mathcal{L}_{quad}$ and the solution to the equations of motion
is, for the following types of fields:
\begin{itemize}
\item Real scalar:
  \begin{eqnarray}
    \mathcal{L}_{quad}&=&-\frac{1}{2}\Phi(D^2 + M^2)\Phi ,\\
    \Phi_c&=&\sum_{n=0}^{\infty}(-1)^n\frac{D^{2n}}{M^{2n+2}}
    \frac{\delta S_{int}}{\delta \Phi}.
  \end{eqnarray}
  
\item Complex scalar:
  \begin{eqnarray}
    \mathcal{L}_{quad}&=&-\Phi^\dagger(D^2 + M^2)\Phi, \\
    \Phi_c&=&\sum_{n=0}^{\infty}(-1)^n\frac{D^{2n}}{M^{2n+2}}
    \frac{\delta S_{int}}{\delta \Phi^\dagger}, \\
    \Phi^\dagger_c&=&\sum_{n=0}^{\infty}(-1)^n\frac{D^{2n}}{M^{2n+2}}
    \frac{\delta S_{int}}{\delta \Phi}.
  \end{eqnarray}
  
\item Real vector:
  \begin{eqnarray}
    \mathcal{L}_{quad}&=&\frac{1}{2}V^\mu
    \left\{\eta_{\mu\nu}(D^2+M^2)-D_\nu D_\mu\right\}V^\nu, \\
     V_c &=& -\frac{1}{M^2}\sum_{n=0}^\infty Q^n
     \frac{\delta S_{int}}{\delta V},
   \end{eqnarray}
   where $Q$ is a differential operator that acts on a Lorentz vector and
   gives a Lorentz vector as:
   \begin{equation}
     (QV)_{\mu}:=\frac{D_\nu D_\mu -\eta_{\mu\nu}D^2}{M^2}V^\nu.
   \end{equation}
   
 \item Complex vector:
  \begin{eqnarray}
    \mathcal{L}_{quad}&=&V^{\dagger\mu}
    \left\{\eta_{\mu\nu}(D^2+M^2)-D_\nu D_\mu\right\}V^\nu, \\
     V_c &=& -\frac{1}{M^2}\sum_{n=0}^\infty Q^n
     \frac{\delta S_{int}}{\delta V^\dagger}, \\
     V^\dagger_c &=& -\frac{1}{M^2}\sum_{n=0}^\infty Q^n
     \frac{\delta S_{int}}{\delta V}.
   \end{eqnarray}
   
 \item Vector-like fermion (using two-component spinor notation):
   \begin{eqnarray}
     \mathcal{L}_{quad} &=& 
     iF^{\dagger}_{L\dot{\alpha}}\bar{\sigma}^{\dot{\alpha}\alpha}_\mu
     D^\mu F_{L\alpha}
     +iF^{\dagger\alpha}_{R}\sigma^\mu_{\alpha\dot{\alpha}}
     D_\mu F^{\dot{\alpha}}_R \nonumber \\
     & & -M(F^\dagger_{L\dot{\alpha}}F^{\dot{\alpha}}_R+
     F^{\dagger\alpha}_RF_{L\alpha}), \\
     (F_c)_{L\alpha} &=&
     \frac{i}{M}\sigma^\mu_{\alpha\dot{\alpha}}D_\mu F_R^{\dot{\alpha}}
     +\frac{1}{M}\frac{\delta S_{int}}{\delta F^{\dagger\alpha}_R}, \\
     (F_c)^{\dot{\alpha}}_R &=&
     \frac{i}{M}\bar{\sigma}^{\dot{\alpha}\alpha}_\mu D^\mu F_{L\alpha}
     +\frac{1}{M}\frac{\delta S_{int}}
     {\delta F^\dagger_{L\dot{\alpha}}}, \\
     (F_c)^\dagger_{L\dot{\alpha}} &=&
     -\frac{i}{M}\sigma^\mu_{\alpha\dot{\alpha}}D^\mu F^{\dagger\alpha}_R
     -\frac{1}{M}\frac{\delta S_{int}}{\delta F^{\dot{\alpha}}_R}, \\
     (F_c)^{\dagger\alpha}_R &=&
     -\frac{i}{M}\bar{\sigma}^{\dot{\alpha}\alpha}_\mu D^\mu
     F^\dagger_{L\dot{\alpha}}
     -\frac{1}{M}\frac{\delta S_{int}}{\delta F_{L\alpha}}.
   \end{eqnarray}
   
 \item Majorana fermion:
   \begin{eqnarray}
     \mathcal{L}_{quad} &=&
     iF^\dagger_{\dot{\alpha}}\bar{\sigma}^{\dot{\alpha}\alpha}_\mu D^\mu
     F_\alpha \nonumber \\ & &
     -\frac{1}{2}\left(\epsilon^{\alpha\beta} F_\beta F_\alpha
       +F^\dagger_{\dot{\alpha}} \epsilon^{\dot{\alpha}\dot{\beta}}
       F^\dagger_{\dot\beta}\right), \\
     (F_c)_\beta &=& \epsilon_{\alpha\beta}
     \left(i \bar{\sigma}_\mu^{\dot{\alpha}\alpha} D^\mu
       F^\dagger _{\dot{\alpha}}
       + \frac{\delta S_{int}}{\delta F_\alpha}\right), \\
     (F_c)^\dagger_{\dot{\beta}} &=& \epsilon_{\dot{\beta}\dot{\alpha}}
     \left(i \bar{\sigma}_\mu^{\dot{\alpha}\alpha} D^\mu F_\alpha
       + \frac{\delta S_{int}}{\delta F^\dagger_{\dot{\alpha}}}\right).
   \end{eqnarray}
   
\end{itemize}

\section{Creation of models}
\label{sec:models}

In this section we will describe how to create a model using the module
\texttt{matchingtools.core}. It assumes that the classes and functions that
are used are in the namespace. To import all the classes and functions that
appear here do
\begin{verbatim}
    from matchingtools.core import (
        Tensor, Operator, OperatorSum
        TensorBuilder, FieldBuilder,
        D, Op, OpSum,
        number_op, power_op)
\end{verbatim}

The \verb|from ... import ...| style is recommended, as the expressions that
appear when using this library tend to be long, so having the short names
directly accessible is preferable.

\subsection{Creation of tensors and fields}

In \emph{MatchingTools}, the basic building blocks for everything are the
objects of the class \texttt{Tensor}, which we simply call tensors here.
Examples of tensors are fields (light and heavy), symmetry group related
tensors (such as Pauli matrices) or coupling constants (including gauge
couplings, Yukawa couplings and masses).

Tensors have an attribute \verb|is_field| that is \texttt{True} if and only if
they are spacetime dependent (i.e., they are fields). Fields can have
derivatives applied to them. The attribute \verb|num_of_der| counts the number
of derivatives that apply to a field. Derivatives are understood here to be
covariant derivatives $D_\mu$ corresponding to the gauge group of the low
energy effective theory. Each derivative applies only to one field. The
Leibniz rule is used whenever a derivative of a product is encountered.
Tensors can be either commuting of anti-commuting, which is distinguished by
the attribute \texttt{statistics}. It can be set equal to either
\texttt{boson} or \texttt{fermion}, both being variables defined in this
module. Finally, all tensors have an attribute \texttt{indices}, a list of
integer numbers representing their tensor indices; and an attribute
\texttt{name}, an identifier. Other attributes, \texttt{content} and
\texttt{exponent}, are for internal use. Names starting with the character
'\texttt{\$}' are also reserved for internal calculations. 

To create the tensors and fields of a model, the classes \verb|TensorBuilder|
and \verb|FieldBuilder| should be used. For example, the Pauli matrices
$\sigma^a_{ij}$ could be defined as
\begin{verbatim}
    sigma = TensorBuilder("sigma")
\end{verbatim}
and then used when needed as \verb|sigma(i1, i2, i3)| where \verb|i1|,
\verb|i2| and \verb|i3| are the indices. Similarly, a boson field $\phi$ (with
its conjugate $\phi^*$) and a fermion $f$ (with its separate chiralities and
their conjugates) are defined as
\begin{verbatim}
    phi = FieldBuilder("phi", 1, boson)
    phic = FieldBuilder("phic", 1, boson)

    fL = FieldBuidler("fL", 1.5, fermion)
    fR = FieldBuidler("fR", 1.5, fermion)
    fLc = FieldBuidler("fLc", 1.5, fermion)
    fRc = FieldBuidler("fRc", 1.5, fermion)
\end{verbatim}

The second argument of \texttt{FieldBuilder} is the dimension of the field.

\subsection{Definition of the interaction Lagrangian}

Once all the tensors are created, we are ready to define the interaction
Lagrangian. It should be a sum of operators, which in turn are just products
of fields. Using the functions \texttt{OpSum} and
\texttt{Op}:

\begin{verbatim}
    int_lag = -OpSum(Op(...), Op(...), ...)
\end{verbatim}

The minus sign is defined for operator sums and individual operators. The
function \texttt{OpSum} creates an object of the class \texttt{OperatorSum}, a
container for a list of operators representing their sum. The function
\texttt{Op} creates an \texttt{Operator} that contains a list of tensors and
represents their product:

\begin{verbatim}
    Op(tensor1(i1, i2, ...), 
       tensor2(i3, i4, ...), ...)
\end{verbatim}

Positive indices are used to express contraction. During the creation of the
model, any index should be contracted with another, so we will only use here
positive ones. When indices are repeated inside the same operator, the
corresponding contraction is understood. For example, the product of tensors
$r_{ij}s_{limnm}t_{njl}$ would be written as

\begin{verbatim}
    Op(r(0, 1), s(3, 0, 4, 5, 4), t(5, 1, 3))
\end{verbatim}

To introduce a covariant derivative inside an operator, the appropriate
function is \texttt{D}, whose first argument is the Lorentz index of the
derivative and whose second one is the tensor to which it is to be applied:

\begin{verbatim}
    D(i1, tensor(i2, ...))
\end{verbatim}

For numeric coefficients, the function \verb|number_op| creates an operator
with only one special tensor representing a number (its name is
\texttt{"\$number"} and has an attribute \texttt{content} with the actual
number). Multiplication is defined for operators, so the operator
$i V_\mu S^*_a D_\mu S_a$ can be expressed as

\begin{verbatim}
    number_op(1j) * Op(V(0), Sc(1), D(0, S(1)))
\end{verbatim}

Tensors representing a symbolic constant exponentiated to some power can be
created using the function \texttt{power\_op}, that takes the base (a string)
and the exponent (a number) (represented by an extra internal attribute of
tensors: \texttt{exponent}) and optionally some indices and returns an
operator containing only the corresponding tensor. This is useful specially
for the masses of the heavy particles, which tend to appear several times with
different powers in all calculations.

A summary of the tools presented in this section is shown in table
\ref{tab:models}.

\begin{table}
  \centering
  \begin{tabular}{r|l}
    Tensors &
    \texttt{t\_name\ =\ TensorBuilder("t\_name")} \\
    Fields &
    \texttt{f\_name\ =\ FieldBuilder("f\_name",\ dim,}\\
    & \hspace{4cm} \texttt{statistics)} \\
    \\
    Lagrangian &
    \texttt{lag\ =\ -OpSum(Op(...),\ Op(...),\ ...)} \\
    \\
    Operators &
    \verb|Op(tensor1(i1, i2, ...), ...)| \\
    Derivatives &
    \verb|Op(..., D(i1, tensor(...)), ...)| \\
    Num. coef. &
    \verb|number_op(number) * Op(...)| \\
    Symb. power &
    \verb|inv_mass_sq = power_op("M", -2)| \\
  \end{tabular}
  \caption{Summary of the tools for the creation of a model.}
  \label{tab:models}
\end{table}

\subsection{Dealing with spinors}

\emph{MatchingTools} uses the two-component spinor formalism to treat spinor
fields following the conventions in \cite{Dreiner:2008tw}. The module
\texttt{matchingtools.core} defines the following tensors to work with them:

\begin{itemize}
\item \texttt{epsUp} and \texttt{epsDown}: the totally anti-symmetric tensors
  $\epsilon^{\alpha\beta}$ and $\epsilon_{\alpha\beta}$ with two undotted
  two-component spinor indices defined by
  $\epsilon^{12} = -\epsilon^{21} = -\epsilon_{12} = \epsilon_{21} = 1$.
\item \texttt{epsUpDot} and \texttt{epsDownDot}: the totally
  anti-symmetric tensors $\epsilon^{\dot{\alpha}\dot{\beta}}$ and
  $\epsilon_{\dot{\alpha}\dot{\beta}}$ with two dotted two-component spinor
  indices given by $\epsilon_{\dot{\alpha}\dot{\beta}} =
  (\epsilon_{\alpha\beta})^*$ and $\epsilon^{\dot{\alpha}\dot{\beta}} =
  (\epsilon^{\alpha\beta})^*$.
\item \texttt{sigma4} and \texttt{sigma4bar}: the tensors
  $\sigma^\mu_{\alpha\dot{\alpha}}$ and
  $\bar{\sigma}_\mu^{\dot{\alpha}\alpha}$ given by
  $\sigma^\mu = (I_{2\times 2}, \vec{\sigma})$ and
  $\bar{\sigma}_\mu = (I_{2\times 2}, -\vec{\sigma})$, where
  $\vec{\sigma}$ is the three-vector of the Pauli matrices.
  The first index of \texttt{sigma4} and \texttt{sigma4bar}
  corresponds to the Lorentz index.
\end{itemize}

\section{Integration}
\label{sec:integration}

This section explains how to use the classes that represent the heavy fields
as well as the function \texttt{integrate}, to integrate them out. They belong
to the module \verb|matchingtools.integration|. To import them do:

\begin{verbatim}
    from matchingtools.integration import (
        RealScalar, ComplexScalar,
        RealVector, ComplexVector,
        VectorLikeFermion, MajoranaFermion,
        integrate)
\end{verbatim}

To integrate out the heavy fields from a previously defined Lagrangian we
should specify which of the fields are heavy. This is done using the classes:

\begin{itemize}
\item \texttt{RealScalar}. Its constructor receives as arguments the name of
  the field and the number of indices it has.
\item \texttt{ComplexScalar}. Requires a field--conjugate field pair. The
  arguments of the constructor are the name of the field, the name of its
  conjugate and its number of indices.
\item \texttt{RealVector}. The arguments are the name of the field and the
  number of indices. The first index of the field is understood to be the
  Lorentz vector index.
\item \texttt{ComplexVector}. The arguments are the name of the field, the
  name of its conjugate and the number of indices. The first index of both
  fields should be their corresponding Lorentz vector index.
\item \texttt{VectorLikeFermion}. The first argument of the constructor is the
  name of the field. The second and third are the names of the left-handed and
  right-handed parts. The fourth and fifth are their conjugates. The last is
  the number of indices. The first index of the each of the four fields is
  taken to be their two-component spinor index.
\item \texttt{MajoranaFermion}. The arguments are the name of the field and
  the name of its conjugate. The first index of both fields should be their
  two-component spinor index.
\end{itemize}

The constructors for the bosons have the optional arguments: \texttt{order}
(default \texttt{2}), specifying the order in $(D/M)^2$ to which the solution
to the equation of motion is to be expanded, and \texttt{max\_dim} (default
4), representing the maximum allowed dimension for the operators appearing in
this expansion. Both bosons and fermions receive the optional argument
\texttt{has\_flavor} (default \texttt{True}) stating whether the heavy field
has a flavor index. In case it is true, the flavor index is taken to be the
last one.

The heavy field classes include the quadratic terms for the kind of particle
they represent, as well as the solutions to the equations of motion presented
in section \ref{sec:theory}. The mass of a field \texttt{f} is represented by
a tensor whose name is of the form \verb|mass = "M" + f.name|. This tensor
has one index if the heavy field has flavor and none otherwise.

Therefore, the first step for integration is defining the heavy fields:

\begin{verbatim}
    heavy_f = HeavyFieldClass("field_name", ...)
\end{verbatim}

Given an interaction Lagrangian \texttt{int\_lag}, the integration is done
using the function \texttt{integrate}, which takes as arguments a list of the
heavy fields, the interaction Lagrangian and a maximum dimension
\texttt{max\_dim} for the operators of the effective theory. It returns the
corresponding effective Lagrangian:

\begin{verbatim}
   heavy_fields = [heavy_f_1, heavy_f_2, ...]
   eff_lag = integrate(
       heavy_fields, int_lag, max_dim)
\end{verbatim}

\section{Transformations of the effective Lagrangian}
\label{sec:transformations}

After integration, the effective Lagrangian contains in general operators that
are not independent. To rewrite it in terms of a set of independent operators
some manipulations are needed, such as using identities for combinations of
tensors related to the symmetry groups, integrating by parts to move
derivatives from some fields to others, or using the equations of motion of
the light fields.

The \texttt{matchingtools.transformations} module introduces the functions for
doing this kind of manipulations and for the simplification of the Lagrangian.
We will describe here the functions that are imported with

\begin{verbatim}
    from matchingtools.transformations import (
        simplify, apply_rules)
\end{verbatim}

First, the function \texttt{simplify} returns a simplified version of the
Lagrangian it gets as an argument. Tensors representing a number that appear
inside an operator are collected and multiplied. Tensors representing a
symbolic constant exponentiated to some power are also collected to give only
one tensor with the correct exponent. \texttt{simplify} also looks for
Kronecker deltas (tensors with the name \texttt{"kdelta"} and two indices)
removes them by contracting the corresponding indices.

The transformations of a Lagrangian are done using what we call here
\emph{rules}. A rule is a pair (a tuple with two elements) whose first element
is an operator representing a \emph{pattern} and whose second element is an
operator sum representing a \emph{replacement}. They are used by the function
\texttt{apply\_rules} to find occurrences of the pattern and replace them by
the replacement. A rule is written as

\begin{verbatim}
    rule = (Op(...), OpSum(Op(...), Op(...), ...))
\end{verbatim}

The indices that appear in tensors inside the rule can be general integer
numbers. Non-negative integers represent contracted indices, as explained in
section \ref{sec:models}. Negative indices are used for free indices and those
in the replacement should match the corresponding ones in the pattern. For
example the substitution of $\sigma^a_{ij}\sigma^b_{kl}$ by
$2\delta_{il}\delta_{kj} - \delta_{ij}\delta_{kl}$ can be done using the rule

\begin{verbatim}
    rule_fierz_SU2 = (
        Op(sigma(0, -1, -2), sigma(0, -3, -4)),
        OpSum(number_op(2) * Op(delta(-1, -4), 
                                delta(-3, -2)),
              -Op(delta(-1, -2), 
                  delta(-3, -4))))
\end{verbatim}

To transform the Lagrangian using integration by parts or equations of motion
of the light fields the user should also specify the corresponding rules following
this procedure.

The function \texttt{apply\_rules} repeatedly tries to apply every rule of a
list to each operator in an operator sum. If the pattern matches some part of
an operator, the rule is applied and the operator sum updated. The first
argument to \texttt{apply\_rules} is the operator sum, the second is the list
of rules and the last one is the number of iterations. It returns the
resulting operator sum.

To rewrite the Lagrangian in terms of a chosen set of independent operators
the procedure is: define the rules to get to the desired basis, add some rules
to identify the operators and apply the function \texttt{apply\_rules}.

The basis operators should be defined using \texttt{tensor\_op}, a function
that creates an operator with one tensor inside whose name is the argument of
the function. Then write a rule to identify it. For example, for the operator
$\mathcal{O}_{\phi D}=(\phi^\dagger D_\mu \phi) (D^\mu \phi)^\dagger \phi$
we would write

\begin{verbatim}
    OphiD = tensor_op("OphiD")
    rule_def_OphiD = (
        Op(phic(0), D(1, phi(0)), 
           D(1, phic(0)), phi(0)),
        OpSum(OphiD))
\end{verbatim}

If the basis operator in question has some flavor indices,
\texttt{flavor\_tensor\_op} is to be used instead of \texttt{tensor\_op}. It
creates a callable object that takes the corresponding free indices as
arguments. As an example, for the operator
$(O_{e\phi})_{ij} = \bar{l}_{L i} \phi e_{R j} \phi^\dagger \phi$
we would have:

\begin{verbatim}
    Oephi = flavor_tensor_op("Oephi")
    rule_def_Oephi = (
        Op(lLc(0, 1, -1), phi(1), eR(0, -2),
           phic(2), phi(2)),
        OpSum(Oephi(-1, -2)))
\end{verbatim}

\section{Output}
\label{sec:output}

The class \texttt{matchingtools.output.Writer} serves to nicely represent an
effective Lagrangian. It is convenient that the final result is represented as
a list of the coefficients of the operators in the basis. That is, if each of
the terms of the Lagrangian contains a tensor that represents an operator of
the basis, we would like to see what are the tensors that multiply each of
them. This is what \texttt{Writer} does. If \texttt{eff\_lag} is our final
effective Lagrangian and \texttt{op\_names} is a list of the names of the
tensors representing the operators in the basis, do

\begin{verbatim}
    eff_lag_writer = Writer(eff_lag, op_names)
\end{verbatim}

The constructor admits an optional argument \texttt{conjugates}, a dictionary
whose keys are the names of all the tensors involved in the final output and
whose values are the names of their conjugates. This helps \texttt{Writer}
collect pairs of conjugate products of tensors returning their real or
imaginary part.

The string representation can be obtained just by using the \texttt{str}
method of the class \texttt{Writer}. To write it to a text file use

\begin{verbatim}
    eff_lag_writer.write_text_file(filename).
\end{verbatim}

The method \texttt{write\_latex\_file} writes a LaTeX file with the
representation. It receives four arguments: the name of the output file, the
LaTeX representation of the tensors, the LaTeX representation of the
coefficients of the basis operators and a list of the strings to be used to
represent the indices. The LaTeX representations are given by dictionaries
whose keys are the names of the tensors to be represented (or whose
coefficient is to be represented) and whose values are the corresponding code.
This code should contain placeholders for the necessary indices written as
\texttt{"\{\}"} (Python's \texttt{format} style). To produce the characters
\texttt{"\{"}, \texttt{"\}"} in the final code they should appear duplicated
in the dictionary values. 

For a better LaTeX output for the numerical coefficients, the
parameter passed to \texttt{number\_op} in the definitions should be either
an \texttt{int} or a \texttt{fractions.Fraction}. In this context, the
imaginary unit can be introduced by multiplying by the operator
\texttt{core.i\_op}.

\section{An example}
\label{sec:example}

In this section we will be creating a simple model to show some of the
features of \emph{MatchingTools}. The model is described as follows: it has
$SU(2)\times U(1)$ gauge symmetry and contains a complex scalar doublet
$\phi$ (the Higgs) with hypercharge $1/2$ and a real scalar triplet $\Xi$ with
zero hypercharge that couple as:

\begin{equation}
   \mathcal{L}_{int} = - \kappa\Xi^a\phi^\dagger\sigma^a\phi
   - \lambda \Xi^a \Xi^a \phi^\dagger\phi,
\end{equation}

where $\kappa$ and $\lambda$ are a coupling constants and $\sigma^a$ are the
Pauli matrices. We will then integrate out the heavy scalar $\Xi$ to obtain an
effective Lagrangian which we will finally write in terms of the operators

\begin{equation}
  \begin{array}{ll}
    \mathcal{O}_{\phi 6}=(\phi^\dagger\phi)^3, &
    \mathcal{O}_{\phi 4}=(\phi^\dagger\phi)^2, \\
    \mathcal{O}^{(1)}_{\phi}= \phi^\dagger\phi 
    (D_\mu \phi)^\dagger D^\mu \phi, &
    \mathcal{O}^{(3)}_{\phi}= (\phi^\dagger D_\mu \phi)
    (D^\mu \phi)^\dagger \phi, \\
    \mathcal{O}_{D \phi} = \phi^\dagger(D_\mu \phi) 
    \phi^\dagger D^\mu\phi, &
    \mathcal{O}^*_{D \phi} = (D_\mu\phi)^\dagger\phi 
    (D^\mu\phi)^\dagger\phi.
  \end{array}
\end{equation}

Notice that this is not an independent set of operators, as some linear
combinations of them are total derivatives. Because the purpose of this
section is to present a very simple model, we will not be doing integration by
parts and therefore we will not simplify the results any further.

\subsection{Creation of the model}

The required imports are

\begin{verbatim}
  from matchingtools.operators import (
    TensorBuilder, FieldBuilder, Op, OpSum,
    number_op, tensor_op, boson, fermion, kdelta)

  from matchingtools.integration import (
    RealScalar, integrate)

  from matchingtools.transformations import  (
    apply_rules)

  from matchingtools.output import Writer
\end{verbatim}

We will need three tensors, the Pauli matrices and the coupling constants:
   
\begin{verbatim}
   sigma = TensorBuilder("sigma")
   kappa = TensorBuilder("kappa")
   lamb = TensorBuilder("lamb")
\end{verbatim}

We will also use three fields: the Higgs doublet, its conjugate and the new
scalar:
   
\begin{verbatim}
   phi = FieldBuilder("phi", 1, boson)
   phic = FieldBuilder("phic", 1, boson)
   Xi = FieldBuilder("Xi", 1, boson)
\end{verbatim}

Now we are ready to write the interaction Lagrangian:
  
\begin{verbatim}
    interaction_Lagrangian = -OpSum(
        Op(kappa(), Xi(0), phic(1), 
           sigma(0, 1, 2), phi(2)),
        Op(lamb(), Xi(0), Xi(0),
           phic(1), phi(1)))
\end{verbatim}

\subsection{Integration}

To integrate out the heavy $\Xi$ we write
  
\begin{verbatim}
  heavy_Xi = RealScalar("Xi", 1, has_flavor=False)
  effective_Lagrangian = integrate(
      [heavy_Xi], interaction_Lagrangian, 6)
\end{verbatim}

\subsection{Transformations of the effective Lagrangian}
After the integration we get operators that contain
$(\phi^\dagger\sigma^a\phi)(\phi^\dagger\sigma^a\phi)$.
This product can be rewritten in terms of the operator $(\phi^\dagger\phi)^2$.
To do this, we can use the $SU(2)$ Fierz identity:

\begin{equation}
  \sigma^a_{ij}\sigma^a_{kl}=2\delta_{il}\delta_{kj}-\delta_{ij}\delta_{kl}.
\end{equation}

We now know that we can define a rule to transform everything that matches the
left-hand side of the equality into the expression in the right-hand side with
the code

\begin{verbatim}
  fierz_rule = (
      Op(sigma(0, -1, -2), sigma(0, -3, -4)),
      OpSum(number_op(2) * Op(kdelta(-1, -4), 
                              kdelta(-3, -2)),
            -Op(kdelta(-1, -2), 
                kdelta(-3, -4))))
\end{verbatim}

We should now define the operators in terms of which we want to express the
effective Lagrangian

\begin{verbatim}
  Ophi6 = tensor_op("Ophi6")
  Ophi4 = tensor_op("Ophi4")
  O1phi = tensor_op("O1phi")
  O3phi = tensor_op("O3phi")
  ODphi = tensor_op("ODphi")
  ODphic = tensor_op("ODphic")
\end{verbatim}
and then use some rules to express them in terms of the fields and tensors
that appear in the effective Lagrangian

\begin{verbatim}
  definition_rules = [
    (Op(phic(0), phi(0), phic(1), phi(1),
        phic(2), phi(2)),
     OpSum(Ophi6)),
    (Op(phic(0), phi(0), phic(1), phi(1)),
     OpSum(Ophi4)),
    (Op(D(2, phic(0)), D(2, phi(0)),
        phic(1), phi(1)),
     OpSum(O1phi)),
    (Op(phic(0), D(2, phi(0)), 
        D(2, phic(1)), phi(1)),
     OpSum(O3phi)),
    (Op(phic(0), D(2, phi(0)), 
        phic(1), D(2, phi(1))),
     OpSum(ODphi)),
    (Op(D(2, phic(0)), phi(0),
        D(2, phic(1)), phi(1)),
     OpSum(ODphic))]
\end{verbatim}

To apply the $SU(2)$ Fierz identity to every operator until we get to the
chosen operators, we do

\begin{verbatim}
  rules = [fierz_rule] + definition_rules
  max_iterations = 2
  transf_eff_lag = apply_rules(
      effective_Lagrangian, rules, 
      max_iterations)
\end{verbatim}

\subsection{Output}

The class \texttt{Writer} can be used to represent the coefficients of the
operators of a Lagrangian as plain text and write them to a file

\begin{verbatim}
  final_coef_names = [
    "Ophi6", "Ophi4", "O1phi",
    "O3phi", "ODphi", "ODphic"]
  eff_lag_writer = Writer(
    transf_eff_lag, final_coef_names)
  eff_lag_writer.write_text_file(
    "simple_example_results.txt")
\end{verbatim}

It can also write a LaTeX file with the representation of these coefficients
and export it to pdf to show it directly. For this to be done, we should
define how the objects that we are using are represented in LaTeX code and the
symbols we want to be used as indices

\begin{verbatim}
  latex_tensor_reps = {"kappa": r"\kappa",
                       "lamb": r"\lambda",
                       "MXi": r"M_{{\Xi}}",
                       "phi": r"\phi_{}",
                       "phic": r"\phi^*_{}"}

  latex_op_reps = {
    "Ophi": 
    r"\frac{{\alpha_{{\phi}}}}{{\Lambda^2}}",
    "Ophi4": 
    r"\alpha_{{\phi 4}}"}
		   
  latex_indices = ["i", "j", "k", "l"]
  
  eff_lag_writer.write_latex(
      "simple_example", latex_tensor_reps, 
      latex_op_reps, latex_indices)
\end{verbatim}

The expected result is a \texttt{.tex} file (ready to be compiled) with the
coefficients of the operators we defined.

\section{Extras for beyond the Standard Model applications}
\label{sec:extras}

\emph{MatchingTools} includes a subpackage called \texttt{extras}, with some
modules defining tensors and rules that are useful for the applications to
physics beyond the Standard Model. These modules are \texttt{SU2},
\texttt{SU3}, \texttt{Lorentz}, \texttt{SM} and \texttt{SM\_dim\_6\_basis}.
Other modules will be added in the future and will be available in the GitHub
repository of the program, as well as in its updates in the pypi repository
\cite{pip}.

\subsection{The \texttt{SU2} module}

This module defines the following tensors related to $SU(2)$:

\begin{itemize}
\item \texttt{epsSU2}: The totally antisymmetric tensor $\epsilon_{ij}$ with
  two doublet indices and $\epsilon_{12} = 1$.
\item \texttt{sigmaSU2}: The Pauli matrices $\sigma^a_{ij}$. The first index is
  the triplet index, whereas the second and third are the doublet ones.
\item \texttt{CSU2} and \texttt{CSU2c}: the Clebsh-Gordan coefficients
  $C^I_{a\beta}$ with the first index $I$ being a quadruplet index, the second
  $a$ a triplet index, and the third $\beta$ a doublet index. The tensor $C$
  contracted with the corresponding three objects produces a singlet.
\item \texttt{epsSU2triplets}: Totally antisymmetric tensor $\epsilon_{abc}$
  with three $SU(2)$ triplet indices such that $\epsilon_{123} = 1$.
\item \texttt{fSU2}: Totally antisymmetric tensor with three $SU(2)$ triplet
  indices given by $f_{abc} = \frac{i}{\sqrt{2}} \epsilon_{abc}$.
\end{itemize}

It also implements the rules for taking expressions with
$\epsilon_{ij}\epsilon_{kl}$, $\sigma^a_{ij}\sigma^a_{kl}$,
$C^I_{ap}\epsilon_{pm}\sigma^a_{ij} C^{I*}_{bq}\epsilon_{qn}\sigma^b_{kl}$
or contractions of anti-symmetric tensors, and rewriting them in terms of
Kronecker deltas. All the rules are collected in the list \texttt{rules\_SU2}.
The LaTeX representation of the tensors defined is given by the dictionary
\texttt{latex\_SU2}.

\subsection{The \texttt{SU3} module}

The $SU(3)$ tensors defined in this module are:

\begin{itemize}
\item \texttt{epsSU3}: Totally antisymmetric tensor $\epsilon_{ABC}$
  with three $SU(3)$ triplet indices such that $\epsilon_{123} = 1$.
\item \texttt{TSU3}: $SU(3)$ generators
  $(T_A)_{BC}  = \frac{1}{2}(\lambda_A)_{BC}$, where $\lambda_A$ are the
  Gell-Mann matrices. The first index is the octet index. The second and third
  are the anti-triplet and triplet ones.
\item \texttt{fSU3}: $SU(3)$ structure constants $f_{ABC}$.
\end{itemize}

The rule for transforming $\epsilon_{ijk}\epsilon_{ilm}$ into a combination of
Kronecker deltas is implemented. It is included in the one-element list
\texttt{rules\_SU3}. The LaTeX representation of the tensors defined is in
\texttt{latex\_SU3}.

\subsection{The \texttt{Lorentz} module}

This module includes the tensors \texttt{epsUp}, \texttt{epsUpDot},
\texttt{epsDown}, \texttt{epsDownDot}, \texttt{sigma4},
\texttt{sigma4bar} from \texttt{matchingtools.operators} and defines:

\begin{itemize}
\item \texttt{eps4}: Totally antisymmetric tensor $\epsilon_{\mu\nu\rho\sigma}$
  with four Lorentz vector indices where $\epsilon_{0123}=1$.
\item \texttt{sigmaTensor}: Lorentz tensor
  \begin{equation}
    \sigma^{\mu\nu}=\frac{i}{4}\left(
      \sigma^\mu_{\alpha\dot{\gamma}}\bar{\sigma}^{\nu\dot{\gamma}\beta}-
      \sigma^\nu_{\alpha\dot{\gamma}}\bar{\sigma}^{\mu\dot{\gamma}\beta}
    \right).
  \end{equation}
\end{itemize}

The list \texttt{rules\_Lorentz} contains the rules for substituting
$\epsilon^{\alpha\beta}\epsilon^{\dot{\alpha}\dot{\beta}}$ by
$\frac{1}{2} \bar{\sigma}^{\mu,\dot{\alpha}\alpha}
\bar{\sigma}^{\dot{\beta}\beta}_\mu$,
$\epsilon_{\alpha\beta}\epsilon_{\dot{\alpha}\dot{\beta}}$ by
$\frac{1}{2} \bar{\sigma}^\mu_{\alpha\dot{\alpha}}
\bar{\sigma}_{\mu,\beta\dot{\beta}}$ and contracted $\epsilon$ tensors
by combinations of Kronecker deltas.

\subsection{The \texttt{SM} module}

Here, the tensors corresponding to the Standard Model fields and its gauge
coupling constants, Yukawa couplings and CKM matrix are defined.

The Standard Model fields are:
\begin{itemize}
\item \texttt{phi} and \texttt{phic}: The Higgs boson and its conjugate. One
  $SU(2)$ doublet index.
\item \texttt{lL} and \texttt{lLc}: The left-handed lepton doublet. Its
  indices are, in order: the two-component spinor index, the $SU(2)$ doublet
  index and the flavor index.
\item \texttt{qL} and \texttt{qLc}: The left-handed quark doublet. Its indices
  are: the two-component spinor index, the $SU(3)$ triplet (or anti-triplet)
  index, the $SU(2)$ doublet index and the flavor index.
\item \texttt{eR} and \texttt{eRc}: The right-handed electron. Indices:
  two-component spinor and flavor.
\item \texttt{uR} and \texttt{uRc}: The right-handed up quark. Indices:
  two-component spinor, $SU(3)$ triplet (or antitriplet) and flavor.
\item \texttt{dR} and \texttt{dRc}: The right-handed down quark. Indices:
  two-component spinor, $SU(3)$ triplet (or antitriplet) and flavor.
\item \texttt{bFS}: $U(1)$ field strength tensor. Two Lorentz vector indices.
\item \texttt{wFS}: $SU(2)$ field strength tensor. Two Lorentz vector indices
  and one $SU(2)$ triplet index.
\item \texttt{gFS}: $SU(3)$ field strength tensor. Two Lorentz vector indices
  and one $SU(3)$ octet index.
\end{itemize}

The constant tensors are:
\begin{itemize}
\item \texttt{gb} and \texttt{gw}: The $U(1)$ and $SU(2)$ gauge coupling
  constants.
\item \texttt{ye}, \texttt{yec}, \texttt{yd}, \texttt{ydc}, \texttt{yu} and
  \texttt{yuc}: The diagonalized Yukawa matrices for the leptons, the down
  quarks, the up quarks and their conjugates. They have two indices: the first
  one corresponds to the flavor of the doublets and the second to the flavor
  of the singlets.
\item \texttt{V} and \texttt{Vc}: CKM matrix.
\end {itemize}

The module also includes a list of rules \texttt{eoms\_SM}, defined to
substitute the equations of motion, replacing derivatives of the Standard
Model fields by a combination of the other fields. There is a dictionary
\texttt{latex\_SM} containing the LaTeX representation of the tensors that are
defined.

\subsection{The \texttt{SM\_dim\_6\_basis} module}

In this module, the basis for the Standard Model effective Lagrangian up to
dimension six that appears in \cite{deBlas:2014mba} is defined. The rules to
identify them are given in the list \texttt{rules\_basis\_definition}. The
LaTeX representation of their coefficients is in \texttt{latex\_basis\_coefs}.
Modules containing other bases, such as the one in \cite{Grzadkowski:2010es},
will be added in the future.

\section{Using \emph{MatchingTools} with other types of fields}
\label{sec:others}

As explained above, \emph{MatchingTools} can integrate scalars, vector-like or
Majorana fermions, and vectors in Lorentz-invariant theories. For this
purpose, several classes representing the heavy fields are supplied. Other
kinds of fields (for instance, with non canonical kinetic terms, spin $> 1$,
or non relativistic) can be treated as well, once the corresponding class for
it is provided.

Specifically, to treat a new type of field one should define a Python class
implementing the following methods:
\begin{itemize}
\item \texttt{equations\_of\_motion}. Receives an \texttt{OperatorSum}
  object representing an interaction Lagrangian. Returns a dictionary whose
  keys are strings with the names of the heavy fields involved (for example, a
  field and its conjugate, if it is a complex boson) and whose values are
  \texttt{OperatorSum} objects representing the corresponding solution to
  their equation of motion. These solutions can be written in terms of other
  heavy fields, but they should be such that iterative substitutions of their
  respective equations motion reaches a point where no heavy fields appear to
  the desired order in the dimension of the operators.
\item \texttt{quadratic\_terms}. Does not have any parameters. Returns the
  kinetic and mass terms of the corresponding heavy field.
\end{itemize}

For the definition of these methods, it is recommended to use the tools
provided by the \texttt{core} module. Once such a class is defined, its
objects can be included in the list of heavy fields to be passed to
\texttt{integration.integrate} and they will be dealt with in the same way as
the others.

\section{Conclusions}
\label{sec:conclusions}

We have presented \emph{MatchingTools}, a Python library implementing symbolic
tree-level integration of heavy fields for any given model. It is also able to
transform the resulting Lagrangian using rules specified by the user to remove
redundant operators. With this program one can safely automatize these kind of
calculations, which practically eliminates the possibility of algebraic errors
and drastically reduces the calculation times. Even calculations with complex
Lagrangians involving $\sim 100$ independent terms (thousands of terms in some
intermediate steps) can be performed in about thirty seconds (using a
$2.6\,\mathrm{GHz}$ Intel Core i5 processor).

\section*{Acknowledgments}
\label{sec:ack}

The author would like to thank J. de Blas, M. P\'erez-Victoria and J. Santiago
for their very useful guidance, comments and corrections.

Funding: This work was supported by the Spanish MECD grant FPU14, the
Spanish MINECO grants FPA2013-47836-C3-2-P and FPA2016-78220-C3-1-P (Fondos
FEDER) and the Junta de Andaluc\'ia grant FQM101.

\bibliographystyle{elsarticle-num}
\bibliography{bibliography}

\end{document}